# Probing the Surfaces of Interstellar Dust Grains: The Adsorption of CO at Bare Grain Surfaces

Helen J. Fraser,[1*] Suzanne E. Bisschop,[1,2] Klaus M. Pontoppidan,[2] Alexander G.G.M. Tielens[3] & Ewine F. van Dishoeck[2]

1. Raymond and Beverly Sackler Laboratory for Astrophysics, Leiden Observatory, Leiden University, Postbus 9513, 2300RA, Leiden, Netherlands

2. Leiden Observatory, Leiden University, Postbus 9513, 2300RA, Leiden, Netherlands

3. The `Kapteyn Institute' of Astronomy, Universiteit Groningen, Postbus 800, 9700 AV Groningen, Netherlands

**ABSTRACT**

A solid-state feature was detected at around 2175 cm$^{-1}$ towards 30 embedded young stellar objects in spectra obtained using the ESO VLT-ISAAC[†]. We present results from laboratory studies of CO adsorbed at the surface of Zeolite wafers, where absorption bands were detected at 2177 and 2168 cm$^{-1}$ (corresponding to CO chemisorbed at the Zeolite surface), and 2130 cm$^{-1}$ (corresponding to CO physisorbed at the Zeolite surface), providing an excellent match to the observational data. We propose that the main carrier of the 2175-band is CO chemisorbed at bare surfaces of dust grains in the interstellar medium. This result provides the first direct evidence that gas-surface interactions do not have to result in the formation of ice mantles on interstellar dust. The strength of the 2175-band is estimated to be $\sim 4 \times 10^{-19}$ cm molecule$^{-1}$. The abundance of CO adsorbed at bare grain surfaces ranges from 0.06 to 0.16 relative to H$_2$O ice, which is, at most, half of the abundance (relative to H$_2$O ice) of CO residing in H$_2$O-dominated ice environments. These findings imply that interstellar grains have a large (catalytically-active) surface area, providing a refuge for interstellar species. Consequently the potential exists for heterogeneous chemistry to occur involving CO molecules in unique surface chemistry pathways not currently considered in gas-grain models of the interstellar medium.

---

# 1. INTRODUCTION

Interstellar dust grains are thought to consist of an amorphous silicate or carbonaceous core surrounded by a molecular ice layer (Draine 2003; Gibb et al. 2004). Surface reactions, on and in the icy mantle, or on the bare grains, are key routes to forming many molecules observed in star-forming regions (van Dishoeck & Blake 1998; Herbst 2000). Presently, little is known about the morphology, or chemistry of these grain surfaces, or the porosity of the grains themselves. From observational evidence of star forming regions and cometary dust, interstellar silicates are most regularly associated with amorphous Olivine or Pyroxene (Henning 2003) i.e. silicates incorporating Mg and Fe in their crystal structure, and studies of pre-solar nebula inclusions in meteoritic samples suggest that these grains may contain embedded metals or sulphides, even in their interstellar state (Bradley et al. 1999).

Spectral features in the 2000 to 2200 $cm^{-1}$ region are characteristically associated with stretching vibrations of doubly and triply bonded molecules. Pontoppidan et al. (2003) have performed an extensive survey in this spectral region with the ESO VLT-ISAAC, of around 40 low- ($< 50\ L_\odot$) and intermediate-mass ($> 50\ L_\odot$) embedded young stellar objects (YSOs). The spectra are dominated by a broad absorption feature, corresponding to condensed CO (CO ice) in pure CO- (2139.9 $cm^{-1}$ and 2143.7 $cm^{-1}$) and $H_2O$- (2136.5 $cm^{-1}$) dominated environments (Tielens et al. 1991), with ro-vibrational transitions of gas-phase CO superimposed, either in absorption or emission (Pontoppidan et al. 2002; Boogert et al. 2002). Analysis of these data revealed a second, weak, interstellar feature in 30 of the objects, as illustrated (for 6 of these objects) in Fig. 1. The band's peak position ranged from 2165 to 2194 $cm^{-1}$, with an average peak-centre of 2173 ± 4 $cm^{-1}$, except in the 'higher' mass stars in the sample (e.g. Reipurth 50) and those low-mass stars located near to the Trapezium O-stars in Orion (e.g. TPSC 78) whose average peak-centre position was 2167 ± 2 $cm^{-1}$ (see Table 4 in Pontoppidan et al. 2003 for full details). Its full width half maximum (FWHM) ranged from 9 to 36 $cm^{-1}$. Henceforth in this paper, this feature is referred to as the 2175-band.

In this article we assign the 2175-band to CO adsorbed directly on the surfaces of bare interstellar grains. In Sect. 2 we address the possible carriers of the 2175-band, drawing on previous literature describing spectral features in this region, and explain why $OCN^-$, dangling Si-H and C-D stretching vibrations were eliminated as potential carriers of the band. Laboratory experiments were conducted under high-vacuum (HV) conditions, where Zeolite wafers were exposed to CO gas, at surface temperatures from 300-100 K, to illustrate that when CO is chemisorbed on specific binding sites at the Zeolite surfaces, characteristic spectral features are observed between 2180 and 2165 $cm^{-1}$ (see Sect. 3 & 4). An excellent fit was obtained between these laboratory spectra and the 2175 $cm^{-1}$ feature in the observations of Pontoppidan et al. (2003) (see Sect. 4). The astrophysical implications of these results are addressed in Sect. 5.

# 2. POTENTIAL CARRIERS OF THE 2175-BAND

The 2175-band is unique. The vibrational modes of CO molecules trapped in low temperature ices invariably range from 2135 to 2155 $cm^{-1}$ (Fraser et al. 2004). In several studies of CO-containing interstellar ice analogues, including for example $CH_4$, $CH_3OH$, $HCOOH$, $CO_2$ or $H_2O$ (Sandford et al. 1988; Schmitt, Greenberg & Grim, 1989; Gerakines et al. 1995; Palumbo 1997; Ehrenfreund et al. 1996, 1997a, b; Collings et al. 2003a,b; Fraser et al. 2004), no CO-ice band has ever been observed as far into the blue as 2175 $cm^{-1}$. In spectra of pure CO ices, a combination band (arising from the fundamental vibrational and translational modes within the solid) appears as a weak, broad feature centred at 2200 cm$^{-1}$, with FWHM of 80-100 cm$^{-1}$ (Ewing & Pimentel 1961), significantly broader than the interstellar feature. To observe this band in the laboratory with transmission infrared spectroscopy requires very thick samples of CO (> 500 monolayers) in which the majority of the CO-ice is crystalline: attempts to fit such laboratory features to the observations were unsuccessful, ruling out the combination band as a feasible carrier. The optical depth of the 2175-band does not scale with that of the CO- or $H_2O$-ice bands in these objects (van Broekhuizen et al. 2004a), although a tenuous correlation was found between the optical depths of the 2175-band and the 2136.5 cm$^{-1}$ component of the CO-ice band (CO trapped in a $H_2O$-dominated environment) (Pontoppidan et al. 2003), suggesting that both features may share the same, or a closely related chemical species.

In spectra of high-mass star forming regions, a band has sometimes been detected at around 2165 cm$^{-1}$ associated with the CN-stretching vibrations of the OCN$^-$ ion (e.g. Pendleton et al. 1999; Hudson, Moore & Gerakines 2001; van Broekhuizen, Keane & Schutte 2004b; Gibb et al. 2004). However, the peak of this 2175-band is clearly shifted from 2165 cm$^{-1}$. Although more difficult to ascertain because of spectral confusion, the optical depth of the 2175-band does not correlate with the optical depth of any 'XCN'-type component of the band (van Broekhuizen et al. 2004a). Even the most blue-shifted laboratory spectra of solid OCN$^-$ reported in the literature peak at only 2170 cm$^{-1}$ (Raunier et al. 2003), as do CN-stretching vibrations in nitriles and isonitriles (Pendleton et al. 1999): alone, all are inadequate for fitting the 2175 cm$^{-1}$ peak position and band profile (van Broekhuizen et al. 2004a). For these reasons, solid OCN$^-$ was eliminated from this study as the primary carrier of the 2175-band.

These deductions, coupled with the detection of the 2175-band in cold, low-mass protostellar sources where ices are not thought to be energetically processed (Langer et al. 2000) suggest that the 2175-band should be attributed to a solid state feature that is distinct from the ice mantles accreted in dense clouds. One carrier candidate could be the dangling Si-H bonds formed at silicate surfaces when they condense in the presence of $H_2$ (Blanco, Fonti & Orofino 1999). Dangling Si-H bonds give rise to very strong, broad, absorption features between 2200 and 2100 cm$^{-1}$, depending on the silicate, which have significantly broader FWHM ($\approx 150$ cm$^{-1}$) than the profile of this 2175-band. Similarly, infrared absorption spectra of hydrogenated amorphous carbon (HAC), particularly deuterated HAC, also exhibit strong, broad features between 2000 and 2250 cm$^{-1}$, associated with C-D and C-H stretching bands (Grishko & Duley 2003). However, once again such features are significantly broader (FWHM $\approx 200$ cm$^{-1}$) than the 2175-band observed here, and would have to be accompanied by additional, even stronger and distinctive features between 1300 and 800 cm$^{-1}$, which have never been detected in the sources mentioned here for which complimentary observations exist (Alexander et al. 2003). For these reasons C-D / C-H and Si-H bonds were also eliminated from this study as potential carriers of the 2175-band.

CO is frequently used as an infrared probe to characterise the surfaces of oxides and zeolites (Hadjivanov & Vayssilov 2002), particularly to identify the active catalytic binding sites on such materials. Zeolites can be thought of as 'exceptional' silicates, with some Si replaced by Al: geologically they are formed by processing of silicates in aqueous environments. Binding sites, corresponding to CO chemisorbed at Brønsted acid-type sites ($Al_2O_3$, Al-OH), Si-OH-Al or at cation inclusions on the surfaces of these materials, have characteristic frequencies between 2160 and 2180 cm$^{-1}$ (Gruver Panov & Fripiat 1996): physisorbed CO is observed at around 2136 cm$^{-1}$ (Hadjivanov & Vayssilov 2002). This does not mean to suggest that Zeolites are present in interstellar space, although they have been identified in meteorites (Woltzka & Wark 1982) but that similar binding sites could exist on interstellar grains, where Al would be replaced by Mg or Fe. The binding sites associated specifically with absorptions at 2175 cm$^{-1}$ correspond to CO adsorbed at cation sites, e.g. $Mg^{2+}$, $Fe^{3+}$, and have also been found on cation-doped silicate surfaces (Hadjivanov & Vayssilov 2002). Since such spectral features are weak and narrow, associated with secondary (physisorption) features at around 2130 cm$^{-1}$, and even require the same carrier species (CO) as the 2136.5 cm$^{-1}$ feature (itself associated with CO trapped in a $H_2O$-dominated matrix), these interactions were adopted as the most likely carrier of the interstellar 2175-band.

## 3. EXPERIMENTAL METHOD

To test the hypothesis that the 2175 cm$^{-1}$ band is related to CO adsorbed directly on 'bare' interstellar grain surfaces, a synthetically pure sample of a naturally occurring Zeolite, caged Clinoptilolite (($K_2$ $Ca_2$ $Na_2$)O-$Al_2O_3$-$10SiO_2$-$8H_2O$), was exposed to CO gas in our laboratory, under pseudo-interstellar conditions. This material was used to 'prove a concept', being readily available, known to possess the binding sites of interest, with a caged structure (i.e. large surface area), and relatively easy to press into a 'pure' wafer of material. 200 μm thick, ≈ 1 cm diameter wafers of pure Clinoptilolite were produced by high-pressure compression of the powdered material. These were then secured onto a specifically designed gold-coated, oxygen-free high-conductivity copper holder, mounted on a He-cooled cold finger, capable of reaching temperatures as low as 10 K. The whole system was positioned at the centre of a high vacuum (HV) chamber, with base pressures of $1 \times 10^{-7}$ mbar, described in detail elsewhere (Gerakines et al. 1995). The sample was baked overnight to over 400 °C to remove its water of crystallisation, then allowed to cool to room temperature, and pumped for a further 24 hours, to ensure the surface did not become re-hydrolysed. The sample was then cooled from 300 to 100 K in a CO atmosphere of a few Torr, to ensure (a) a constant dynamic monolayer coverage of CO at the Zeolite surface, (b) maximum occupancy of the chemisorption sites, and (c) that any vacuum contaminants (especially $H_2O$), which may preferentially occupy or freeze-out on the wafer surface under rare vacuum conditions, were displaced. 0.5 cm$^{-1}$ resolution Fourier Transform Infrared (FTIR) Transmission Spectra were recorded prior and during CO exposure, at surface temperatures from 300 – 100 K. The surface temperature was never lowered beyond this value to avoid CO condensation to form CO-ice, which occurs at surface temperatures of around 30 K under these experimental conditions (Fraser et al. 2004). Although the Zeolite wafer is transparent in the 2200 - 2000 cm$^{-1}$ region, spectra of the wafer prior to CO exposure were subtracted from the experimental data to ensure that any spurious or temperature effects from the Zeolite spectrum did not influence the final spectrum.

## 4. RESULTS

### 4.1 Laboratory Spectra

Three absorption features, related to CO adsorbed at the surface of the Zeolite wafer, were observed from 300 – 100 K. The 100 K spectrum is shown in Fig. 2(a). Sub-structure and line-centres of the absorption feature components were obtained from a derivative plot of the spectrum. Gaussian components at 2177, 2168, and 2130 cm$^{-1}$ (plus three minor components at 1870, 1887, and 1900 cm$^{-1}$) were required to make a Levenberg-Marquardt non-linear least squares regression fit to the absorption spectrum (dark-grey trace Fig. 2(a)). The strongest of these component features were assigned, by comparison with reported CO vibrational frequencies on Zeolite surfaces (Datka et al. 1999), to physisorbed CO (2130 cm$^{-1}$), and chemisorbed-CO, at either Brønsted acid type binding sites (2168 cm$^{-1}$), or a combination of the Si-OH-Al and cation binding sites (2177 cm$^{-1}$) (see Table 1). Illustrations of these binding sites, and their locations within or on the Zeolite structure are shown in Fig. 2(b). No obvious distinction could be made between the spectra of the chemisorbed bands recorded at different temperatures, although the weak physisorbed CO feature shifted from 2120 to 2130 cm$^{-1}$ as the surface temperature was lowered. As expected, no 'pure' CO-ice features (around 2139 cm$^{-1}$) were observed.

**4.2 Estimating the Band Strength of the 2175 cm$^{-1}$ Laboratory Feature**

In the absence of published absorption coefficients for CO adsorbed on Zeolite, the results from Sect. 4.1 were used to estimate the occupancy of CO molecules at the binding sites on the wafer surface, assuming that a dynamic monolayer of CO was fully covering all the available (internal and external) surface area – a reasonable assumption under these experimental conditions. The total surface area of each wafer was estimated to be around 500 cm$^2$, given that the specific surface area of Clinoptilolite is 20 – 30 m$^2$ g$^{-1}$ (Yasyerli et al. 2002) and the mass of one wafer was approximately 2 mg. It is important to note that this *total* surface area is significantly greater than the *geometric* surface area of the wafer itself ($\approx$ 1 cm$^2$). Assuming a binding-site density of $1 \times 10^{15}$ binding-sites cm$^{-2}$, this equates to a total of 5 ($\pm$ 1) $\times 10^{17}$ binding sites over the whole wafer surface, which will be distributed according to the stoichiometry of the Zeolite, i.e. 12.6 % Brønsted-acid type binding sites (Al-OH, Al$_2$O$_3$), 64.7 % Silicate-OH binding sites (Si-OH-Al) and 22.7 % cation sites (see Table 1). Assuming only one CO molecule was adsorbed at each binding site, and that the absorption coefficient of CO at each of the chemisorption sites is approximately equivalent, (again a reasonable assumption as there is very little difference in the vibrational frequencies of CO adsorbed at the different sites, indicating that there is almost no change in the strength of the C≡O bond), one would expect the integrated absorbencies of the fitted components in the absorption spectrum to track the relative abundances of the binding sites at the wafer surface, i.e. around 7 times as many (cation + silicate) binding sites on the surface in comparison to Brønsted acid-type binding sites. However, as can be seen from the fitted components shown in Fig. 2(a), the integrated intensity of the 2168 cm$^{-1}$ fitted component is almost equivalent to that of the 2177 cm$^{-1}$ component. Furthermore, some CO is observed to be physisorbed to the surface, so it is likely that not all the chemisorption sites are populated. Therefore, assuming that the Al$_2$O$_3$ and Al-OH binding sites are fully occupied, but that the cation and Si-OH-Al sites are not (fully occupied), it was estimated that over the total surface area of the wafer $\approx 1.6 \times 10^{17}$ CO molecules were occupying surface chemisorption sites and contributing to the laboratory spectrum, as detailed in Table 1. This corresponds to a chemisorbed CO density of $3.2 \times 10^{14}$ molecules cm$^{-2}$, with a band strength A of $\sim 4 \times 10^{-19}$ cm molecule$^{-1}$, given that the optical depth of the laboratory spectrum at 2175 cm$^{-1}$ is $\sim$ 0.06. This number, uncertain by a factor of about 3, is consistent with the concept that the band is weak, but accounts for a significant population of CO molecules on a geometrically small wafer, which due to its morphology has a very large total surface area.

Due to fact that this experiment was conducted in a CO-gas environment, the remainder of the binding sites at the surface (of the whole wafer) were occupied by a dynamic monolayer of physisorbed CO molecules evinced by the (weak) 2130 cm$^{-1}$ component absorption feature in Fig. 2, and possibly a few chemisorbed H$_2$O molecules that were not desorbed during the wafer preparation (see Sect. 3). At least as many, and probably more CO molecules would have been physisorbed ($\leq 3.4 \times 10^{17}$) than chemisorbed ($\approx 1.6 \times 10^{17}$) to the wafer surface, illustrating, as expected, that the absorption coefficient of physisorbed CO is significantly less than that of chemisorbed CO. Under these experimental conditions the lifetime of physisorbed CO at the wafer surface will be less than 20 ns: consequently if the CO-gas environment is removed the physisorption band will also disappear. In interstellar regions, at grain temperatures above the condensation temperature of CO, a CO molecule may occupy a physisorption site, but never for long enough that a significant physisorbed population would amass, so the 2120-2130 cm$^{-1}$ band will never be observed. As is already well established in the literature, below the condensation temperature of CO, CO-ice may form, with its characteristic vibration at around 2139 cm$^{-1}$. No further discussion of physisorbed CO is made in this paper.

### 4.3 Comparison with Observations

Fig. 1 shows detailed comparisons between the laboratory CO-Zeolite spectra (grey dashed line) and six of the YSOs observed by Pontoppidan et al. (2003) (black line). In certain objects, such as IRS63, HH100 IR and Elias 32 (Fig. 1 (c) (d) and (f) respectively), only the CO-Zeolite spectrum makes a significant contribution to the 2175-band, and the laboratory spectrum can be scaled to provide an excellent match to the observational profile. However, in many cases it seems a second carrier contributes to the 2175 cm$^{-1}$ profile, so each spectrum was subsequently fitted (grey solid line) by a combination of the laboratory CO-Zeolite spectrum (from Figure 2), plus a Gaussian fixed at a centre position of 2165 cm$^{-1}$ and with FWHM of 25 cm$^{-1}$ (grey dotted line), representative of laboratory OCN$^-$ spectra (Hudson et al. 2001; van Broekhuizen et al. 2004b). For convenience, the intense CO ice band was excluded from the fit. Details of the fitting of this band and a combined fit to the 2175 cm$^{-1}$ and CO-ice features can be found elsewhere (Pontoppidan et al. 2003; van Broekhuizen et al. 2004a). As can be seen from Fig. 1, no OCN$^-$ contribution was required to fit the 2175-band profile in the low-mass sources, e.g. IRS63, HH100 IR or Elias 32. Along lines of sight that are heated or possibly subject to strong ultraviolet radiation from nearby objects (e.g. Reipurth 50 Fig. 1(a) and TPSC 78 Fig. 1(b)), the OCN$^-$ feature contributes more significantly to the overall profile, and in these cases the weak interstellar feature peaks closer to 2165 cm$^{-1}$. In Fig. 1(e), EC90a is an example of an object where the CO gas-grain surface interaction and OCN$^-$ feature contribute almost equally to the overall band profile.

## 5. ASTROPHYSICAL IMPLICATIONS

### 5.1 The Adsorbed CO Abundance

The column density associated with CO adsorbed directly at grain surfaces along any one observational line of site (henceforth called CO gas-grain or CO$_{gg}$), was estimated (within a factor of about 3) from the optical depth at 2175 cm$^{-1}$ of the scaled laboratory CO$_{gg}$ spectrum required to fit the observational data, assuming a band strength (+ estimated error) equivalent to that derived in Sect. 4.2. These data, shown in Table 2, range from 1.1 to 3.2 × 10$^{17}$ cm$^{-2}$, suggesting that about twice as many CO molecules are found in water-rich ice environments compared to those found on bare-grain surfaces. In Table 2 the abundances (relative to H$_2$O-ice) of pure CO-ice, CO in a H$_2$O-dominated ice matrix, and CO$_{gg}$ are also compared. The relative abundance of CO$_{gg}$ is reasonably constant between all the YSOs. No relationship is apparent between the relative abundances of $CO_{gg}$ and pure CO or $H_2O$ ice. This lack of scaling with ice column density is addressed further in Sect. 5.2.

CO may adsorb at bare grain surfaces arising for example, (a) where ices have not yet accreted or formed (see Fig. 3 (a)), or (b) where CO molecules migrate through the ice layer and adsorb to the underlying grain surface (see Fig. 3 (b)). Depending on the nature of the binding sites, CO desorption enthalpies on Zeolite surfaces are reported to range from ≈ 2400 to ≈ 12,990 K (Hadjivanov & Vayssilov 2002), with the majority of CO desorbing from sites with desorption enthalpies above ≈ 6000 K. The reported binding energies for CO adsorbed at $Mg^{2+}$ and Si-OH-Al type binding sites on doped Zeolite surfaces are ≈ 4930 – 10,850 K and ≈ 5530 - 7500 K respectively (He et al. 1992; Hadjivanov & Vayssilov 2002). These values are significantly higher than the CO-CO binding energy in CO ices (860 K) or the CO-$H_2O$ binding energy in binary ice systems (1180 K) (Collings et al. 2003a,b), and (at the lowest limits) comparable with the desorption enthalpy of $H_2O$ ice (5600 K (amorphous) to 5570 K (crystalline) (Fraser et al. 2001)).

Similar behaviour to (b) has been observed in haloform-ice-Platinum systems (Greca et al. 2004). The haloform molecules diffuse 'through' pores and cracks in $H_2O$ ice, and on reaching the interface between the ice and the supporting substrate chemisorb to the substrate rather than remaining bonded to the ice surface, as the haloform-substrate bond is an energetically more favourable state. From laboratory work (Collings et al. 2003a,b) and observational data (Pontoppidan et al. 2003), it is clear that CO ice adsorbed above $H_2O$-ice can subsequently diffuse and become trapped within pores in the $H_2O$-ice matrix. In the laboratory the process is initiated by heating, analogous to thermal processing in protostellar regions, but as the energy barriers to "diffusion of CO into $H_2O$-ice pores" and "CO desorption from CO dominated ice layers" are almost identical (Collings et al. 2003b), it is also possible, given the timescales available in interstellar regions, that CO could equally migrate through the porous structure of interstellar $H_2O$ ices and reach the ice-grain interface at grain temperatures as low as 15 K. This scenario may additionally help to explain why Pontoppidan et al. (2003) observed a weak correlation between the 2175 $cm^{-1}$ and 2136.5 $cm^{-1}$ features in the analyses of their observational spectra, since both features rely on the migration of CO molecules within the porous structure of $H_2O$ ice.

A schematic representation of the different (postulated) ways a $CO_{gg}$ population could accumulate at interstellar grain surfaces is shown in Fig 3. Initially, a $CO_{gg}$ population is accumulated in regions where ices are just starting to (or have not yet) formed, and surface formation of $H_2$ and the gas-phase formation of CO (from $C^+$ + OH) are key molecular processes (see Fig. 3(a) i). From simple abundance arguments, $CO_{gg}$ accretion must compete with other 'bare-grain' processes, e.g. $H_2$ formation and $H_2O$ formation, so it is unlikely that the whole grain surface would ever become saturated by a monolayer of only $CO_{gg}$ (see Fig. 3(a) ii). As a consequence of some CO molecules being chemisorbed at the grain surface, formation of $CO_2$ (CO + O) and HCO (H + CO) may also be feasible (see Sect 5.3). Eventually, ice mantles, dominated by $H_2O$, will accumulate on the grain surface. $CO_{gg}$ acquired prior to ice formation will be more tightly bound to the grain surface than the ice layer itself (see Sect 5.3) and therefore become buried under the ice layer as it grows (see Fig. 3(a) iii and iv). CO may still be accreted from the gas phase, but from energetics arguments would still be more likely to seek a chemisorption binding site at the remaining regions of bare grain surface than on the ice surface itself. As the ice becomes "thicker" the whole grain will appear to be covered by a mantle, although due to the porous nature of amorphous $H_2O$-ice, the molecular solid would only be tethered to the grain surface at a few points, leaving much of the surface 'bare' even though incoming molecules accreting at the ice surface can no longer "see" the grain (see Fig. 3(a) iv).

Fig. 3(b) illustrates the coldest densest interstellar regions, where $H_2O$-dominated ices have already formed and CO is frozen-out, forming CO ice: some $CO_{gg}$ may already be buried under the ice layer, accumulated in direct gas-grain interactions (see v). Over the timescales available in interstellar regions or in regions where the grains are even mildly heated (15 < grain 'temperature' < 40 K), some of the CO will migrate along the surfaces of pores in the $H_2O$-ice. Since the pore structure is 3 dimensional and interconnected, some pores will inevitably lead from the $H_2O$-CO ice interface to the $H_2O$-ice – grain interface. CO can not migrate through the $H_2O$-ice bulk. As some of the grain surface is still 'bare', at least some of the migrating CO molecules could preferentially adsorb at the grain surface rather than the ice surface (see Fig. 3(b) vi). This process continues until all the binding sites at the grain surface are occupied, or no more CO molecules can reach the bare grain surface (see Fig. 3(b) vii). The $CO_{gg}$ population becomes saturated. In this instance the optical depth of the $CO_{gg}$ band would also become saturated as first observed by Pontoppidan et al. (2003). Such suggestions are entirely consistent with the fact that a $CO_{gg}$ component is also observed on "warmer" lines of sight to high-mass YSOs (van Broekhuizen et al. 2004a), where the $CO_{gg}$ feature exists, but is overwhelmed by the nearby, stronger $OCN^-$ band, peaking at 2165 $cm^{-1}$, arising from chemical processing of the ice mantle. Finally, as the binding energy of $CO_{gg}$ is much greater the desorption enthalpy of $H_2O$ ice (see Sect. 5.3), $CO_{gg}$ molecules remain bound to the grain surface during, and even after ice desorption (for example in hot cores and close to the star in protostellar regions). Consequently, CO is available to react with radicals, ions, and larger molecules that may have been formed in the ice, and may even provide a carbon repository to react with transient species from the gas phase in subsequent stages of stellar evolution.

Spectra of other molecules adsorbed on Zeolites are well documented in the literature (Bordiga et al. 2000), as are the catalytic properties of such surfaces (Weitkamp 2000). In the interstellar case, $H_2O$ is one other obvious candidate that could also chemisorb at these catalytic binding sites, besides CO. In an extensive search of the literature, it was not possible to find any measurements of $H_2O$ binding energies at Zeolite surfaces; possibly because surface analysis of such materials is conducted using CO, and requires the water-of-crystallisation to be removed. However, to "dry" Zeolite crystals, temperatures in excess of 650 K are required, indicating that the majority of $H_2O$ molecules must desorb at energies in excess of ≈ 20,000 K. In interstellar regions, the difficulty in detecting $H_2O$ adsorbates lies in distinguishing the surface hydroxyl (Si-OH) and water-hydroxyl (H-OH) stretching vibrations from each other, and from any $H_2O$ ice bands in the same line of sight, since all three sets of vibrations lie between 3500 and 3000 $cm^{-1}$. The only potential band that might be observed is a sharp, narrow, intense feature at around 3620 $cm^{-1}$, related to the stretching vibration of a subset of the surface hydroxyl groups, which increases in intensity and broadens if a strong hydrogen bond forms between the surface and the adsorbate, with the adsorbate acting as the acceptor molecule (Hadjiivanov & Vayssilvov 2002).

## 5.2. Surface Area of Interstellar Dust

Assuming that all grains are chemically equivalent in these different objects, one may at first suspect that the $CO_{gg}$ column density should scale with dust column density (or $H_2$ column density) and CO-gas column density. $H_2$ column densities are not available for the objects discussed here, but are typically a factor of $10^4$ to $10^5$ greater than that of the $H_2O$-ice column density in YSOs (e.g. Whittet et al. 2001; Pontoppidan et al. 2004). However, no clear scaling pattern emerges as illustrated in Table 2. The average abundance of $CO_{gg}$ adsorbed directly on the grain-surface binding sites is $\approx 2.4 \times 10^{-6}$ per H-nucleus. Assuming only one CO molecule occupies each chemisorption site and that the occupied surface area per binding site is approximately 30 $\text{Å}^2$, the total surface area occupied by $CO_{gg}$ on interstellar grains is at least 7 ($\pm$ 3) $\times 10^{-21}$ cm$^2$ per H-nucleus, and possibly much larger if only a fraction of the catalytically-active surface sites are occupied. The uncertainty in this figure is derived from a combination of the factor of 3 uncertainty in band strength derived from the estimated occupancy of the surface binding sites in the experiment, and an additional maximum estimated error of 20 per cent in the $CO_{gg}$ abundance per H-nucleus, which arises from assuming that $n(H_2O)/n(H_2)$ is $\approx 9 \times 10^{-4}$ on all the lines of sight presented here. For comparison, the total surface area of interstellar silicates has previously been estimated as 1 - 2 $\times 10^{-21}$ cm$^2$ per H-nuclei, assuming that the grains are solid spheres (e.g. Mathis, Rumpl & Nordsieck 1977; Walmsley, Flower & Pineau des Forêts 2004).

Many authors characterise interstellar grains using the concept of porosity, defined as the ratio between the solid volume of the grain to the total volume of the grain. Indeed, Mathis (1998) has suggested that the observed strength of the interstellar silicate band is only reproducible if interstellar silicate grains are formed from amorphous silicates, containing almost 100% of the Si, Fe, and Mg abundance, with > 25 % of their total volume being "vacuum" (Mathis & Whiffen 1989). Porous grains have both internal and external surface areas, which would be accessible if CO molecules were first adsorbed on the external grain surface, then migrated into the pores. This could account for the difference between the two estimates of total surface area (of interstellar grains) per H-nucleus given above. However porosity is only a measure of the volume-filling factor of a grain and says little of its morphology: the collision cross section of a porous grain hardly differs in comparison to a solid grain containing the same mass of material, so the probability of CO molecules colliding with the grains would be much less than if the same mass were contained in fractal grains, whose open and filamentary structures exhibit increases in both collision cross section and surface area when compared to solid grains of the same mass (Fogel & Leung 1998). The large interstellar grain surface area derived above from the $CO_{gg}$ interaction, coupled with the $CO_{gg}$ column density, seem compelling evidence that interstellar grains deviate significantly from "solid spheres" and are most likely porous, irregularly shaped aggregates with very large surface areas.

Consequently, the $CO_{gg}$ abundance traces the total grain surface area, i.e. the morphology of interstellar grains. Assuming interstellar grains are porous and fractal, the $CO_{gg}$ abundance will not just depend on CO-gas – interstellar grain collision rates, but also the number of available binding sites on the grains in the line of sight, which may depend on diffusion rates and geometric or chemical accessibility to the binding sites, factors related to the dust morphology rather than its abundance. As grains aggregate and grow larger, the total grain surface area will reduce, and the $CO_{gg}$ abundance will go down. Where gas-phase CO is adsorbed on bare grains, CO gas-phase abundances (2-3 $\times 10^{-4}$ relative to $H_2$ (Lacy et al. 1994)) could be depleted by around 5 %, even in regions where no CO ice is detected. At an extinction $A_J$ = 1 mag (roughly corresponding to an $A_v \approx$ 3 - 4 mag, depending on the extinction law applied), this depletion corresponds to an optical depth of the $CO_{gg}$ band of $\approx$ 0.013, assuming $N_H = 5.6 \times 10^{21}$ cm$^{-2}$ mag$^{-1}$ $A_J$ (Vuong et al. 2003). It is therefore possible that $CO_{gg}$ bands could be detected above the noise limit (i.e. at optical depths $\approx 0.02$) from lines of sight where $A_J \geq 1.5$ (roughly corresponding to an $A_v \approx 4.5 - 6$ mag). This is above the critical extinction threshold for $H_2O$ ice detection in Taurus (Whittet et al. 2001), but below the $A_J$ at which CO-ice could be detected. It will be an exciting prospect to find a line of sight where the weak 2175 cm$^{-1}$ feature is observed but CO-ice is not.

**5.3 Heterogeneous Chemistry**

The potential for heterogeneous chemistry to occur at catalytically-active surfaces on interstellar grains opens up new pathways towards molecular complexity that could be as important in the solid-state chemistry of star-forming regions as the chemistry occurring at the ice surface or in the ice film itself. Put simply, heterogeneous catalysis requires the surface binding site to facilitate the surface reaction, participating in the reactions without being incorporated in the product. Effectively, the presence of a catalyst always lowers the energy barrier to a reaction, either by acting as a donor or acceptor of electrons, weakening the molecular bond and forming a transition state complex, or sterically orientating the reagents, so specific products are formed. The more effective the catalyst, the more significant the effects. For example, Woon and co-workers have used quantum chemistry to show that in the presence of $H_2O$ ice, the barriers to H-atom addition and abstraction reactions may be lowered by between $\approx$ 700 and 1900 K, e.g. $H + H_2CO \rightarrow CH_3O$ (Woon 2002), and that ionisation barriers are reduced by up to 50 per cent when certain species are solvated in $H_2O$ ice, e.g. $H_2$, $CH_3OH$ (Woon 2004). Silicon and metal-cations are very effective catalysts since they have a range of accessible unoccupied atomic orbitals, making electron transfer and transition-state complex formation particularly simple. Consequently, silicon-based and cation-doped surfaces will make better catalysts than $H_2O$, although in the absence of any comparative reaction studies in the literature this is difficult to prove unequivocally.

In 'non-icy' environments, it is already known that $H_2O$ and $H_2$ form on bare grains from their atomic constituents (O'Neil & Williams 1999; Herbst 2001). Similar reactions, involving $CO_{gg}$, could account for the high abundances of $CH_3OH$ and $CO_2$ in interstellar ices, which may be difficult to explain by ice chemistry alone (Tielens & Hagen 1982; van der Tak, van Dishoeck & Caselli 2000; Watanabe, Shiraki & Kouchi 2003). On cation doped silica surfaces, more representative of the interstellar case, CO-binding energies have been reported to range from around $\approx$ 4930 K to 8900 K (Xu & Goodman 1993), dominated by CO desorption from the most strongly bound sites. Assuming the adsorption and desorption enthalpies of these systems are similar, these figures imply that CO could become chemisorbed at bare grain surfaces prior to ice formation. In the absence of grain heating or sufficient energy to break the grain-CO bond, CO would then be likely to participate in Langmuir-Hinshelwood or Eley Rideal-type surface reactions, most likely involving an adsorbed or 'incoming' O or H atom, leading to HCO and $CO_2$ formation (Fig. 3(a)). New reaction pathways may also be opened up, for example, if the chemisorbed CO becomes trapped under the ice layer when it forms, the $CO_{gg}$ may react with ice constituents, e.g. free radicals or atoms, or may recombine with larger molecules forming heavier organics as the ice layer subsequently desorbs (Fig. 3(b)). Alternatively, as recent experimental work has shown, photochemistry of the grain – metal cation – adsorbate – ice system could generate a plethora of new molecules in the solid state (Gleeson et al. 2003; Bergeld et al. 2004). Consequently, the "high-temperature" desorption of chemisorbed CO from bare-grain surfaces will have an important impact on the organic inventory of warm protostellar regions, retaining a C-source in the solid state to much higher temperatures than previously assumed, and possibly even beyond the desorption threshold of the ice layers themselves. Finally, chemisorbed CO will participate in surface reactions at higher temperatures when compared to the same reactions occurring on or in ice mantles, which may well have influenced the organic inventory of the early solar system (Pearson et al. 2002).

The adsorption of other molecular species, such as $N_2$, to the same surface binding sites may lead to molecular dissociation, which activates otherwise chemically inert molecules, providing a direct route towards the formation of bio-molecular precursors, some of which have been tentatively detected in interstellar regions, e.g. Glycine (Kuan et al. 2003). Furthermore, reactions involving heavy isotopes are known to be enhanced at Zeolite surfaces (Lee at al. 1999), the heavier isotopes desorbing at higher temperatures, leading to isotope enhancement effects, (e.g. $^{13}C$ and D containing molecules).

## 6. CONCLUDING REMARKS

Prior to this work, no direct evidence existed for gas-surface dynamics in the interstellar medium. It was assumed that CO-gas and interstellar dust collide and interact, but only such that CO molecules might condense and form CO-ice. The experimental data presented in this paper corroborate our initial assumption that the (weak) 2175 $cm^{-1}$ feature is likely to be associated with CO molecules chemisorbed to bare interstellar grain surfaces. This identification suggests that gas and grains can interact without forming ice, and that CO, as well as other atomic and molecular species, may be adsorbed directly at bare interstellar grain surfaces. This observation has immediate implications, firstly as to how much CO gas might be depleted onto grain surfaces, apparently invisible in gas-phase or ice spectra, secondly impacting on our understanding of the chemical and physical morphology of interstellar grain surfaces, and finally in terms of the subsequent heterogeneous chemistry that may result from CO being chemisorbed, rather than frozen-out, on grain surfaces. These deductions open up a new set of discussions on the synergy between gas and dust in star forming regions, the nature of interstellar dust, and the chemistry occurring there.

## ACKNOWLEDGMENTS

The authors wish to acknowledge a NWO SPINOZA grant and NOVA, the Dutch Research School for Astronomy, for funding this work, and an award of large programme observing time on the ESO-VLT-ISAAC facilities at Paranal, Chile, where the observations in this article were made. We are grateful to Prof. Pierre Jacobs of the Catholic University of Leuven, Belgium, for his assistance in producing the wafers for this experiment, and to Alsipenta, Germany and St. Clouds Mining Company, USA, who supplied the Zeolite materials for this study free of charge.

**FIGURE CAPTIONS**

Figure 1: Detailed fits to 6 spectra of young stellar objects where the unidentified 2175 cm$^{-1}$ band was observed with the ESO VLT-ISAAC (Pontoppidan et al. 2003), a) Reipurth 50, b) TPSC78, c) IRS63, d) HH100 IR, e) EC 90A, f) Elias 32. For clarity, the scale is set to truncate the intense pure CO ice feature at 2139 cm$^{-1}$ and this band is excluded from the fit. The overall fit (grey solid line) comprises the laboratory data shown in Fig. 2 (light grey dashed line), and a Gaussian component, fixed at 2165 cm$^{-1}$, representing the "XCN" feature, attributed to the presence of OCN$^-$ in interstellar ice (light grey dotted line). Observational data are shown in black; all observations show a combination of broad solid state and sharp narrow gas-phase CO features. Fits to Elias 32 and HH 100 IR appear to fit poorly in the 2152 cm$^{-1}$ region due to the presence of the bright Pfβ hydrogen recombination line at 4.687 μm.

Figure 2: (a) Laboratory spectrum of CO adsorbed on the internal and external surfaces of a Clinoptilolite wafer (solid black line) at 100 K. Sub-structure and line-centres of the absorption feature components were obtained from a derivative plot of the spectrum (not shown). A Levenberg-Marquardt (L-M) non-linear least squares regression fit to the spectrum, is shown as a solid grey line: the main peak comprises three components at ~ 2177 cm$^{-1}$, (dash-dotted line; assigned to CO chemisorption at cation and Si-OH-Al sites), 2168 cm$^{-1}$ (dashed line; assigned to CO chemisorption at Al-OH / Al$_2$O$_3$ sites) and 2130 cm$^{-1}$ (dotted line; assigned to CO physisorption): weak features are also observed at 1870, 1887 and 1900 cm$^{-1}$. (b) The locations of various CO binding environments within and on the caged structure of Clinoptilolite are indicated, and the chemical structure of each binding site shown schematically.

Figure 3: A schematic representation of the processes leading to the accumulation of CO molecules adsorbed directly at interstellar grain surfaces. (a) CO$_{gg}$ accreted directly from the gas-phase, in regions where ices are just starting to (or have not yet) formed. (b) CO$_{gg}$ formed indirectly, in regions where CO and H$_2$O ice mantles have amassed and CO migrates to the grain surface. See text for a more detailed discussion.

Table 1. Distribution of the available binding sites in caged Clinoptilolite versus their occupation in these experiments.

| Binding Site | $\nu(CO)_{ads}$ $(cm^{-1})^a$ | % of sites on wafer surface | No. of occupied surface sites in these experiments |
|---|---|---|---|
| Gas* | 2143 | - | - |
| Ice* | 2139 | - | - |
| Al-OH, $Al_2O_3$ | 2169 | 12.6 | $6.3 \times 10^{16}$ (2168 $cm^{-1}$ component) |
| Si-OH-Al | 2175 | 64.7 | |
| $Na^+$ | 2178 | | $9.5 \times 10^{16}$ |
| $K^+$ | 2180 | 22.7 | (2177 $cm^{-1}$ component) |
| $Ca^{2+}$ | 2177 | | |
| $^\dagger Mg^{2+}$ | 2178 | - | - |
| $^\dagger Fe^{2+}$ | 2180 | - | - |

* Included for comparison only

$^a$ Dakta et al. (1999)

$^\dagger$ Not relevant to this Zeolite but included for interstellar relevance

Table 2. Comparison of CO column densities in ice versus 'bare'-grain environments.

| Source | $CO_{gg}$ | | Pure CO-ice | | CO in $H_2O$-rich ice | |
|---|---|---|---|---|---|---|
| | column density $\times 10^{17}$ cm$^{-2}$ | relative abundance (to $H_2O$)[‡] | column density $\times 10^{17}$ cm$^{-2}$ | relative abundance (to $H_2O$)[‡] | column density $\times 10^{17}$ cm$^{-2}$ | relative abundance (to $H_2O$)[‡] |
| Reipurth 50 | 3.2 | 0.06 | 3.4 | 0.07 | 7.3 | 0.14 |
| TPSC 78 | 2.1 | 0.08 | 0.8 | 0.03 | 3.7 | 0.14 |
| IRS 63 | 2.6 | 0.16 | 11.2 | 0.68 | 4.9 | 0.29 |
| HH100IR | 2.9 | 0.12 | 8.8 | 0.37 | 4.4 | 0.18 |
| EC 90A | 1.1 | 0.08 | 13.9 | 1.07 | 4.4 | 0.34 |
| Elias 32 | 2.4 | 0.15 | 11.7 | 0.75 | 6.9 | 0.44 |

[‡] $H_2O$ column densities taken from van Broekhuizen et al. (2004a), using

$$N = 300 \times \tau(3.07\mu m)/A_{H_2O}$$

where $A_{H_2O} = 2.0 \times 10^{-16}$ cm molecule$^{-1}$ (d'Hendecourt & Allamandola 1986)

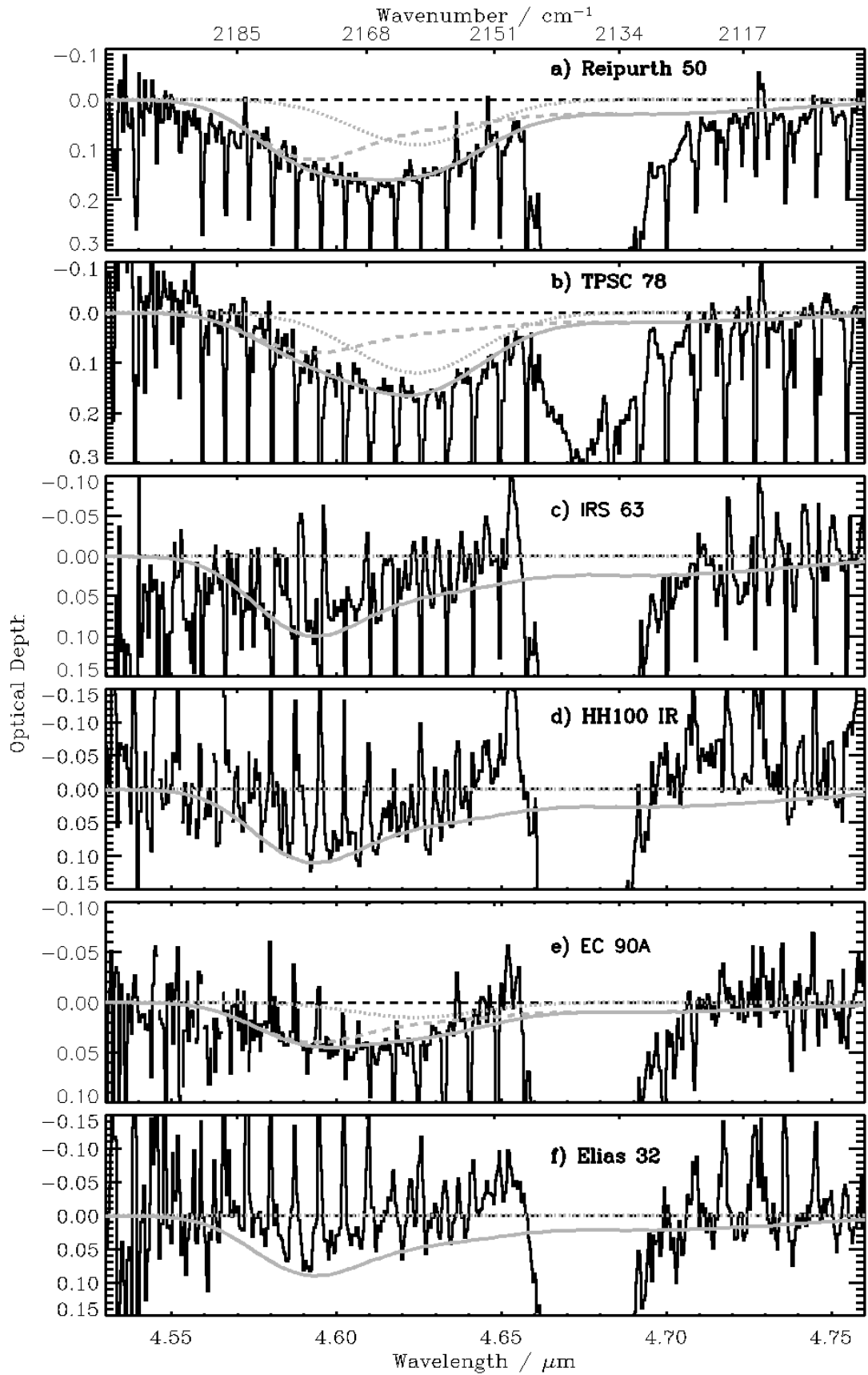

Figure 1. Fraser et al. Probing the Surfaces of Interstellar Dust Grains

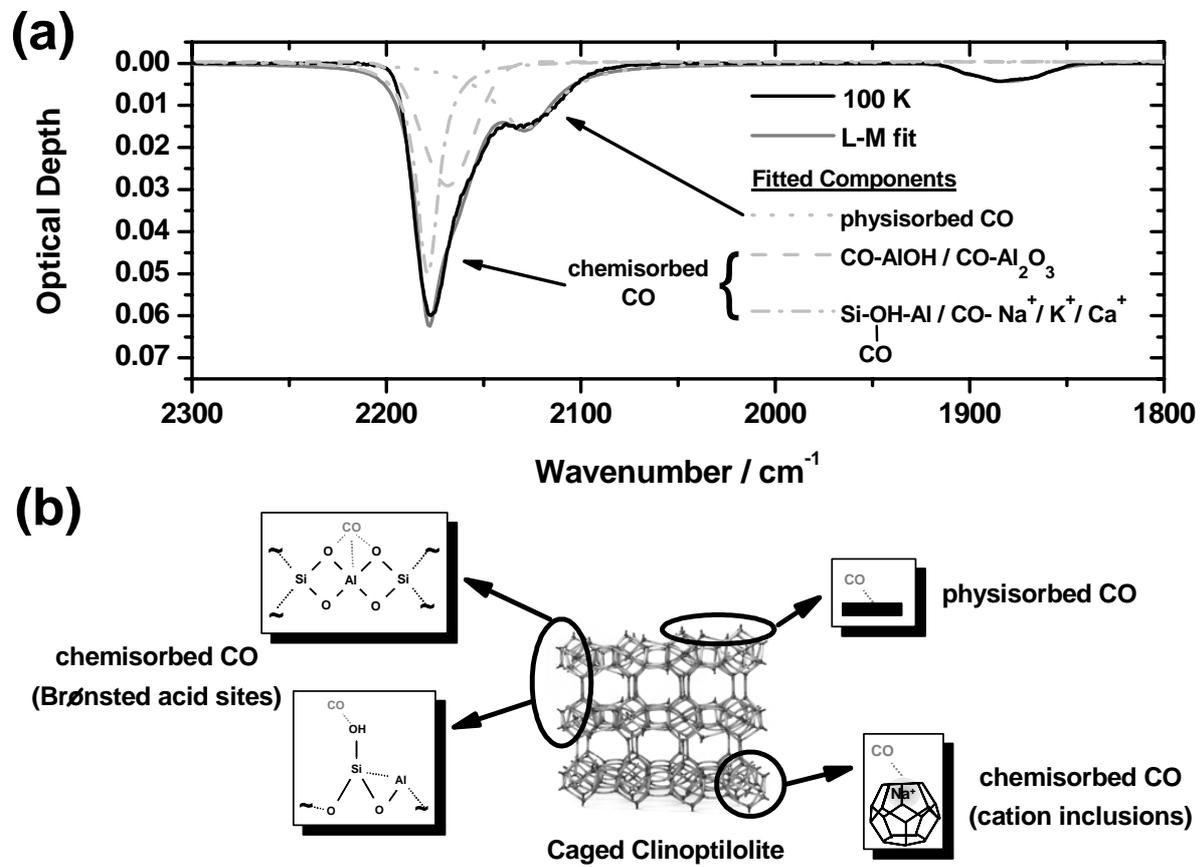

Figure 2. Fraser et al. Probing the Surfaces of Interstellar Dust Grains

Figure 3. Fraser et al. Probing the Surfaces of Interstellar Dust Grains

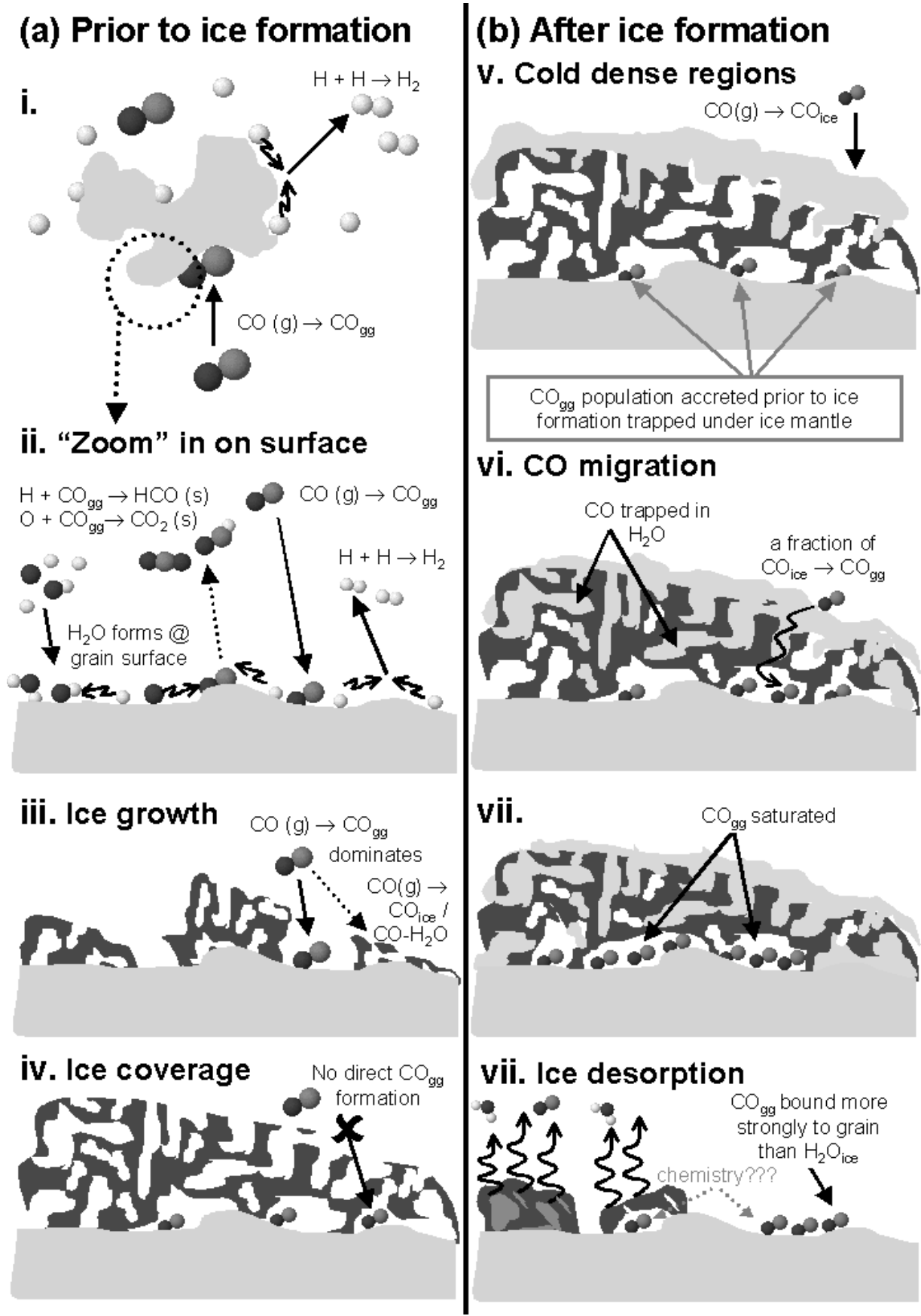